\documentclass[letterpaper, 10 pt, conference]{ieeeconf}  

\IEEEoverridecommandlockouts                              

\usepackage{xcolor}
\usepackage{latexsym,amsmath,amssymb,amsfonts,mathrsfs,mathtools}
\usepackage{url}
\usepackage{comment}
\usepackage{cite}

\usepackage{epsfig} 
\usepackage[colorlinks=true, linkcolor=black, citecolor=black, urlcolor=blue]{hyperref}
\hypersetup{breaklinks=true}

\title{\LARGE \bf
Trial-Level Time-frequency EEG Desynchronization as a Neural Marker of Pain
}

\author{D.A. Blanco-Mora$^{1}$, A. Dierolf$^{2}$, J. Gon\c{c}alves$^{1}$, and M. van Der Meulen$^{2}$ 
\thanks{DAB is supported by the FNR project INTER/DFG/21/15020234/BIML-19.}
\thanks{$^{1}$ Luxembourg Centre for Systems Biomedicine, University of Luxembourg, Luxembourg}%
\thanks{$^{2}$ Department of Behavioural and Cognitive Sciences, University of Luxembourg, Luxembourg}%
\thanks{{\small Emails:} {\tt\small \{diego.blanco-mora, angelika.dierolf, jorge.goncalves, marian.vandermeulen\}@uni.lu}}
}

\begin{document}

\maketitle
\thispagestyle{empty}
\pagestyle{empty}

\begin{abstract}

Pain remains one of the most pressing health challenges, yet its measurement still relies heavily on self-report, limiting monitoring in non-communicative patients and hindering translational research. Neural oscillations recorded with electroencephalography (EEG) provide a promising avenue for identifying reproducible markers of nociceptive processing. Prior studies have reported pain-related event-related desynchronization (ERD) in the alpha and beta bands, but most rely on trial-averaging, obscuring variability that may be critical for perception.
We analyzed high-density EEG from 59 healthy participants who underwent electrical stimulation under Pain and No-Pain conditions. Per-trial time–frequency decomposition revealed robust beta-band ERD in frontal-central electrodes that differentiated Pain from No-Pain trials. Generalized linear mixed models demonstrated that ERD scaled with subjective intensity ratings (VAS), and that age and gender moderated this relationship. Reverse models further showed that ERD predicted VAS ratings across participants, underscoring its potential as a nonverbal marker of pain.
These findings provide preliminary evidence that trial-level EEG oscillations can serve as reliable indicators of pain and open avenues for individualized, report-free pain monitoring. Future work should validate these results in patient populations and extend analyses to multimodal approaches combining EEG, MRI, and attention-based modulation strategies.

\end{abstract}

\section{Introduction}

Pain is one of the most pervasive health problems worldwide, affecting quality of life, productivity, and healthcare systems. Despite its enormous personal and societal costs, pain assessment remains largely dependent on subjective self-report. While self-report measures such as the visual analogue scale (VAS) provide a straightforward and widely adopted approach, they are limited by their reliance on verbal or behavioral communication, and by their inherent subjectivity. This poses challenges in contexts where patients are unable to report their pain accurately, such as in infants, patients with cognitive impairments, or sedated individuals. It also constrains the ability to monitor pain dynamically in experimental and clinical research. These limitations have motivated a search for objective neurophysiological markers of pain, capable of providing reproducible indicators that complement or even substitute self-reports \cite{Ploner2017}.

Electroencephalography (EEG) has emerged as a promising tool in this endeavor. With millisecond temporal resolution and relative affordability, EEG enables the study of oscillatory brain dynamics that are sensitive to nociceptive input. Pain-related EEG responses are typically characterized by changes in rhythmic activity, most prominently event-related desynchronization (ERD) in the alpha (8--12 Hz) and beta (13--30 Hz) frequency bands, which reflect reductions in synchronous neural firing following sensory input \cite{Peng2014, Nickel2017, Hu2019}. Across studies, painful stimulation has been shown to elicit reproducible alpha and beta ERD in sensory and frontal cortices, supporting their role as candidate neural markers of pain \cite{Dowman2008}. Conversely, increases in oscillatory power (event-related synchronization, ERS) have been linked to top-down modulation of pain through attention and expectation \cite{Diers2020, Ecsy2018}. This dual pattern suggests that oscillatory activity not only encodes nociceptive input but also reflects cognitive control processes shaping pain perception.  

Although these findings highlights the potential of oscillatory EEG markers, several gaps remain. First, most existing studies rely on trial averaging to extract oscillatory responses, which improves signal-to-noise but loses trial-to-trial variability that may be critical for capturing perceptual fluctuations \cite{Ploner2017}. Single-trial approaches are needed to quantify how neural oscillations dynamically relate to pain perception on a moment-to-moment basis. Second, the relationship between oscillatory markers and subjective intensity ratings remains poorly defined. While some studies have shown correlations between beta/alpha desynchronization and perceived pain, the predictive strength and specificity of these associations remain unclear \cite{Nickel2017}. Third, little is known about how demographic factors such as age and gender shape oscillatory pain responses. Previous work suggests that pain modulation mechanisms decline with age, and that gender differences in pain perception may be mediated by neurophysiological differences \cite{vanderMeulen2024}. Yet, few EEG studies have systematically incorporated these factors into predictive models.  

Recent methodological advances enable progress on these challenges. Generalized linear mixed models (GLMMs), for example, allow for modeling trial-level data while accounting for participant-specific variability, making them well suited to evaluate oscillatory markers in pain research \cite{Hu2019}. Furthermore, emerging evidence shows that attention-driven interventions, such as sustained gaze fixation or rhythmic visual flicker, can externally induce synchronization in frontal or occipital networks and modulate pain-related responses \cite{Veniero2021, Quentin2016, Ecsy2018, LopezDiaz2021}. These findings highlight the translational potential of oscillatory markers not only for pain quantification but also for developing novel modulation strategies.  

In the present study, we provide preliminary evidence that frontal-central EEG oscillations can serve as trial-level indicators of pain. Using high-density EEG recordings from 59 healthy participants exposed to electrical Pain and No-Pain conditions, we examined event-related desynchronization (ERD) in the alpha and beta bands. Time--frequency features were extracted with a well-established data-driven approach that identifies functionally relevant oscillatory patterns without imposing a priori constraints. This methodology has previously been applied in domains such as motor coordination and aging \cite{vanHoornweder2022a, vanHoornweder2022b}, as well as in studies of transcranial electrical stimulation \cite{BlancoMora2024, vanHoornweder2025}. Our analyses tested whether ERD distinguishes pain conditions, scales with subjective intensity ratings, and interacts with demographic factors such as age and gender. By applying generalized linear mixed models (GLMMs) and reverse prediction models, we demonstrate that ERD can predict subjective ratings while capturing individual variability, supporting its utility as a nonverbal marker of pain. These findings address key gaps in the literature by moving beyond trial averaging, incorporating contextual and demographic variables, and directly linking oscillatory markers to subjective intensity. Together, this work lays the foundation for individualized, trial-level pain decoding and opens avenues for multimodal and intervention-based approaches to pain neuroscience.

\section{Methodology}

\subsection{Participants}
Participants were drawn from the dataset originally described in \cite{Rischer2025}. In that study, ninety-five healthy adults were recruited, including younger (18--36 years) and older participants (60--82 years). Recruitment was carried out at the University of Luxembourg and in the wider community via advertisements, media coverage, and outreach in senior centers. Exclusion criteria included a diagnosis of chronic pain, neurological or psychiatric disorders, substantial cognitive impairment (Mini-Mental State Examination score $\leq 24$), and the use of medication affecting pain perception or cognition on the day of testing. Ethical approval was granted by the University of Luxembourg Ethics Review Panel, and all participants provided written informed consent in accordance with the Declaration of Helsinki.

The final sample size in the original study was 39 younger and 42 older adults \cite{Rischer2025}. In the present work, a subset of the original 81 participants included in the parent study was analyzed, usable EEG datasets were available for 59 participants due to technical issues and data quality constraints. This reduced but balanced sample comprising 59 participants in total, 29 Older Adults (17 Males, 12 Females) and 30 young adults (13 Males, 17 Females) provided sufficient statistical power for the present analyses. 

\subsection{Experimental Paradigm}
The experimental paradigm was based on the protocol reported by \cite{Rischer2025}, with adaptations for the present analyses. Painful electrical stimulation was delivered using a concentric surface electrode (Brainbox Ltd., Cardiff, UK) attached to the volar surface of the left forearm. Each stimulus consisted of a 500 ms train of biphasic square-wave pulses at 100 Hz, generated by a custom-built constant current stimulator. 

Electrical intensities for the \textit{Pain} and \textit{No-Pain} conditions were individually calibrated during a preliminary calibration phase. Target values on a 100-point visual analogue scale (VAS) were approximately 62.5 for the Pain condition (moderate pain) and below the pain threshold for the No-Pain condition (VAS $< 25$). This ensured that stimulation levels were adapted to each participant’s subjective sensitivity. 

During the acquisition phase, stimuli were presented in blocks of five consecutive trials. Blocks alternated between Pain and No-Pain conditions, resulting in a balanced design of 25 trials per condition per participant. Each trial began with an auditory cue (500 ms) followed by an anticipation interval (3500 ms), the 500 ms stimulation, and a response period during which participants rated pain intensity and unpleasantness on the VAS.

\subsection{EEG Acquisition}
EEG data were acquired using a 64-channel Brain Products system (Brain Products GmbH, Gilching, Germany) with Ag/AgCl electrodes positioned according to the international 10--20 system. Signals were sampled at 1000 Hz, with FCz as the online reference and AFz as ground. Four additional electrodes recorded vertical and horizontal electro-oculograms, and auxiliary channels recorded electrocardiogram and electromyogram activity. Electrode impedance was maintained below 10 k$\Omega$. Data were recorded in an electrically shielded, light-attenuated chamber.

\subsection{EEG Preprocessing}
Preprocessing was carried out in MATLAB (2021a, The MathWorks Inc., Portola Valley, CA, USA) using EEGLAB \cite{Delorme2004} and in-house scripts. Continuous EEG was re-referenced to the common average and down-sampled to 250 Hz. Band-pass filtering (1--40 Hz, zero-phase FIR filter \cite{Gustafsson1996}) and a 50 Hz notch filter were applied. Noisy channels were detected and removed using the CleanRawData plugin and then interpolated using spherical splines.

Independent Component Analysis (ICA) was run with runica on data concatenated across conditions and with PCA rank set to \((\textit{nbchan} - \textit{elim\_chans} - 1)\). When auxiliary (non-EEG) channels were present, ICA was computed on the EEG-only subset, and the resulting unmixing/sphere matrices were transferred back to the full dataset. Artefactual components were identified using a \emph{hybrid} procedure: (i) probabilistic ICLabel classification \cite{PionTonachini2019} with parameterized thresholds, and (ii) correlation with auxiliary channels. For ICLabel, components with ``Brain'' probability $< 0.5$ or with probability $\geq 0.3$ in non-brain classes (Eye, Muscle, Heart, Line Noise, Channel Noise, Other) were flagged. In parallel, correlations were computed between IC activations and the following auxiliary channels: vertical electrooculogram (VEOG), horizontal electrooculogram (HEOG), electrocardiogram (EKG), and electromyogram (EMG). Components showing an absolute Pearson correlation $> 0.4$ with any of these channels were also flagged. Components identified by either criterion were removed using \texttt{pop\_subcomp}. Finally, a surface Laplacian transform was applied to reduce volume conduction and enhance spatial specificity.

\subsection{Time--Frequency Analysis}
Time--frequency (TF) decomposition was performed with complex Morlet wavelet convolution \cite{Cohen2014}. For each electrode of interest (F1, F2, Fz, FCz), trials were analyzed over 8--30~Hz using 100 logarithmically spaced frequencies. Wavelets were defined on a symmetric time window centered on zero with fixed 10-cycle width across frequencies, and convolution was implemented in the frequency domain (FFT), using power-of-two zero padding for efficiency. The per-trial complex analytic signals were obtained by inverse FFT and trimmed to account for wavelet edge effects (see below).

Baseline correction used a pre-stimulus interval of $-500$ to $-200$~ms. We computed ERD/ERS as percentage change relative to baseline:
\[
\mathrm{ERD/ERS}(t,f) = 100 \times \left(\frac{P(t,f)}{\overline{P}_{\text{base}}(f)} - 1\right),
\]

Here $P(t,f)$ denotes instantaneous power and $\overline{P}_{\text{base}}(f)$ the mean baseline power. Following the implementation, the baseline term used for single-trial ERD/ERS was the \emph{grand} (trial-averaged) baseline power at each frequency to ensure consistency between grand-average and trial-wise transforms.

To mitigate edge artifacts, we used two safeguards implemented in code: (i) an \emph{optional} trial-wise temporal padding mode that replicates the signal at both ends before convolution, with padding length equal to the wavelet window half-width; and (ii) post-convolution temporal cropping that removes the final portion of the TF matrix approximately equal to three-quarters of the wavelet length. The retained time axis therefore spans from the baseline onset ($-500$~ms) to the cropped post-stimulus analysis endpoint ($1000$~ms), ensuring minimal contamination by convolution transients.

\begin{figure*}
  \centering
  \includegraphics[width=0.75\linewidth]{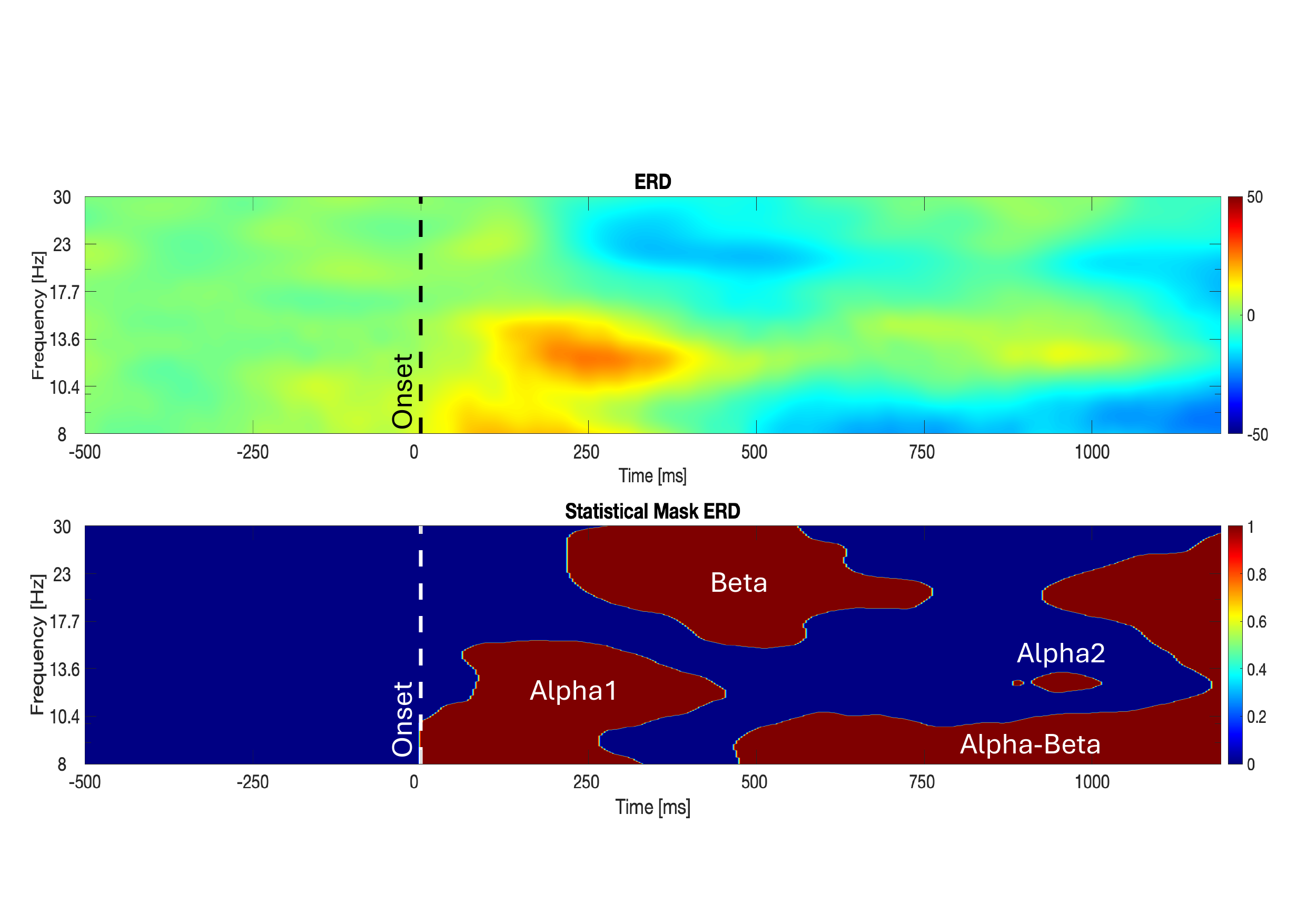}
  \caption{Time-Frequency feature extraction. Top: Grand Time-frequency Average across participants and conditions; Bottom: Significance mask for the found Regions Of Interest (ROI).}
  \label{fig:TF_steps}
\end{figure*}

\paragraph{Data-driven TF masking.}
We derived task-relevant TF regions using a data-driven masking procedure applied to a grand-average TF matrix computed across participants and conditions (blinded to all factors). The grand-average was restricted to $-500$ to $1000$~ms to further limit edge effects. For each TF bin within 0--1000~ms, we compared post-stimulus power against the baseline distribution ($-500$ to $-200$~ms) via pointwise $t$-tests and controlled family-wise error across all TF comparisons. The corrected significance threshold was $\alpha \sim 10^{-6}$. Contiguous significant clusters were identified and labeled as \textit{Alpha1}, \textit{Alpha2}, \textit{Alpha-Beta} and \textit{Beta} regions of interest (ROIs), see it in \ref{fig:TF_steps}. Specifically, time--frequency decompositions were first computed separately for each of the selected frontal--central electrodes (F1, F2, Fz, FCz). For every trial, ERD/ERS values were then averaged across these electrodes, then, ERD/ERS values were then extracted by averaging within each ROI (frequency × time) and carried forward to the statistical models. Unless otherwise specified, references to ``higher ERD/ERS'' denote larger absolute percentage-change magnitudes.

\paragraph{Outputs for modelling.}
In addition to the condition-wise grand-average TF matrices, we retained single-trial TF power arrays of shape [frequency $\times$ time $\times$ trials] to enable per-trial statistical analyses (GLMMs) and to compute ROI-averaged ERD/ERS on a trial-by-trial basis.

\subsection{Statistical Modelling}
All statistical analyses were performed in RStudio (version 2024.12.1) using the \texttt{lme4} package \cite{Bates2015, RStudio2020}. Generalized linear mixed models (GLMMs) were employed with a Gamma distribution and log link, chosen to accommodate the strictly positive distribution of transformed ERD values. Participant-level random intercepts were included to account for interindividual variability. 

We first modeled ERD as a function of experimental condition (Pain vs. No-Pain), age group (Young vs. Old), and gender (Male vs. Female), including all multi-way interactions between fixed factors. A stepwise backward elimination strategy was applied to obtain parsimonious models while preserving explanatory power. To avoid redundancy, in this paper we exposed the results of the stepwise backward reduced model.  

Second, reverse models were also constructed to predict VAS from ERD and contextual factors, thereby evaluating the feasibility of EEG oscillatory features as nonverbal predictors of pain intensity. For all analyses, a two-sided significance threshold of $\alpha = 0.05$ was adopted. Model fits were compared using likelihood ratio tests, Akaike information criterion (AIC), and Bayesian information criterion (BIC). Residual distributions, dispersion statistics, and outlier diagnostics were systematically inspected to ensure model validity.  

\subsection{Summary}
This methodological framework combines high-density EEG acquisition, rigorous preprocessing and artifact rejection, trial-level time--frequency analysis, and mixed-effects statistical modeling. By integrating subjective ratings and demographic variables into the models, our aim is to establish EEG-derived ERD as a reproducible and individualized marker of pain processing.  

\begin{figure*}
  \centering
  \includegraphics[width=.49\linewidth]{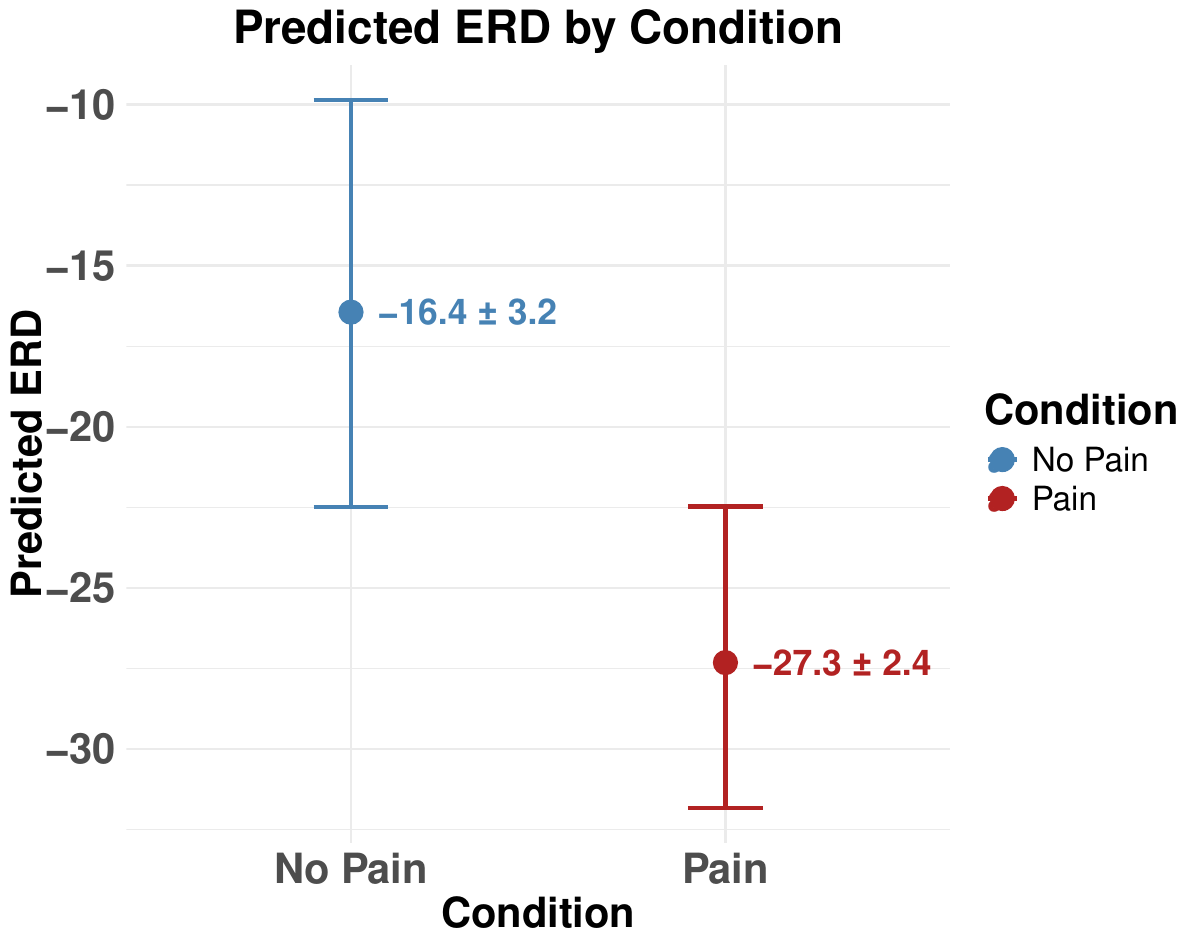}
  \includegraphics[width=0.49\linewidth]{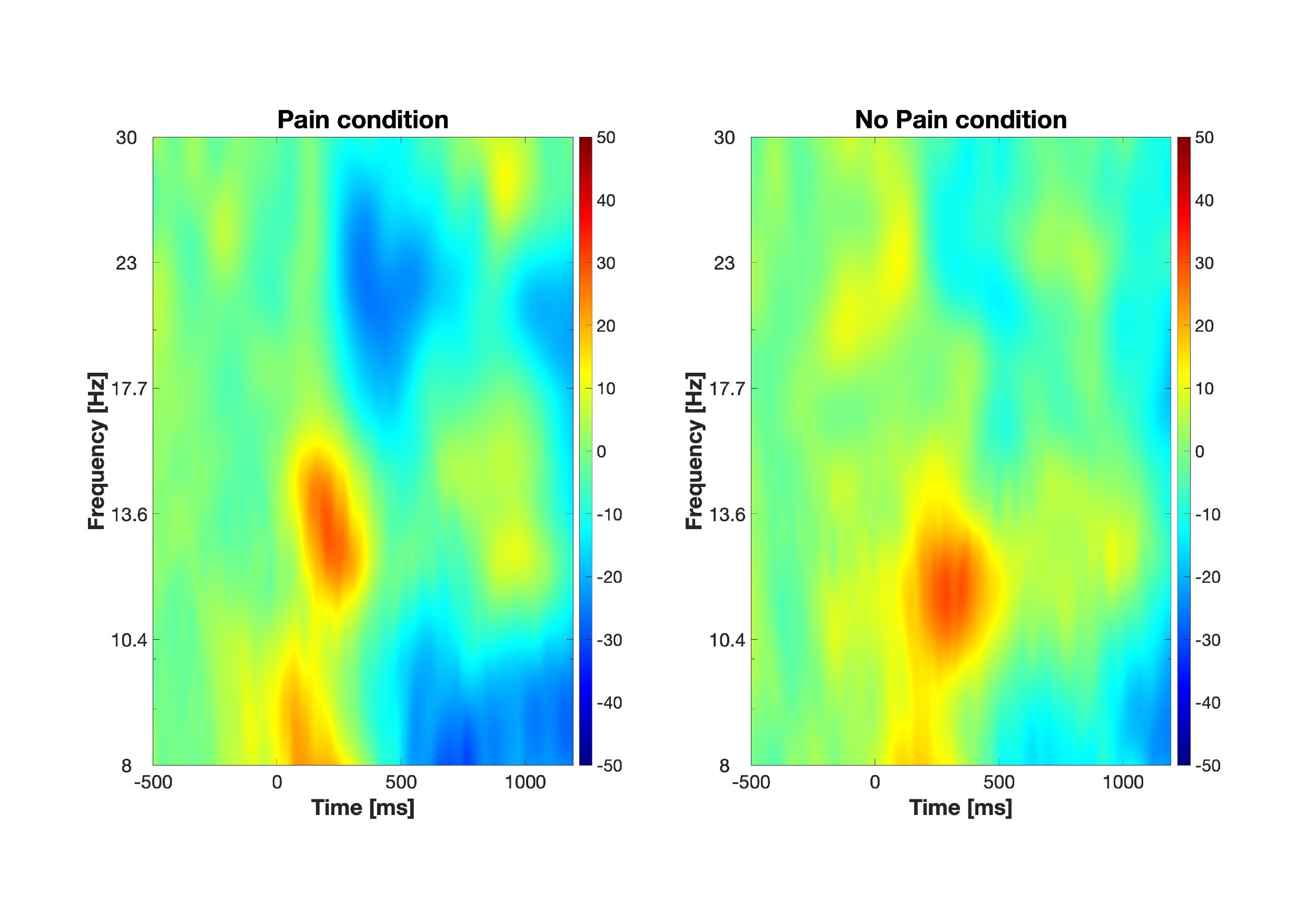}
  \includegraphics[width=.49\linewidth]{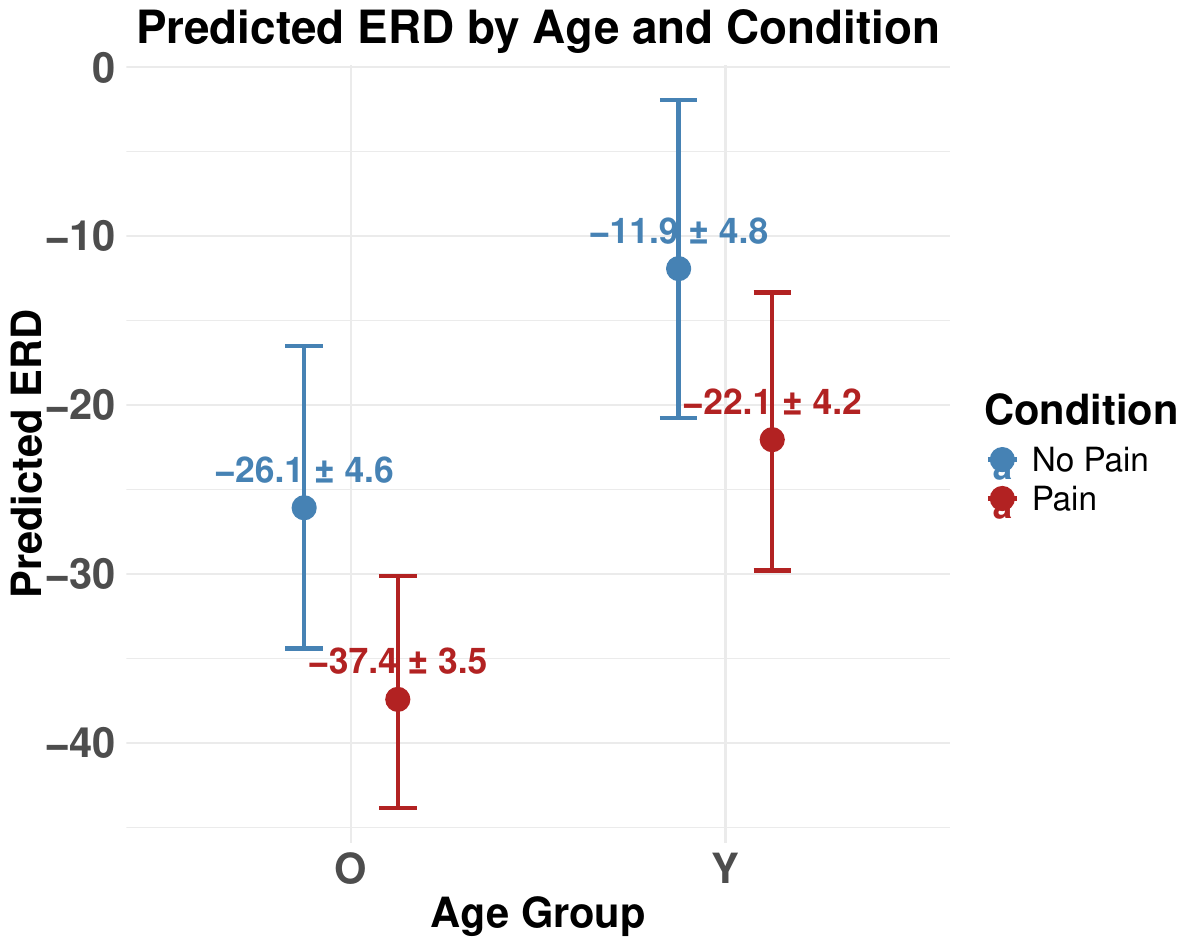}
  \includegraphics[width=.49\linewidth]{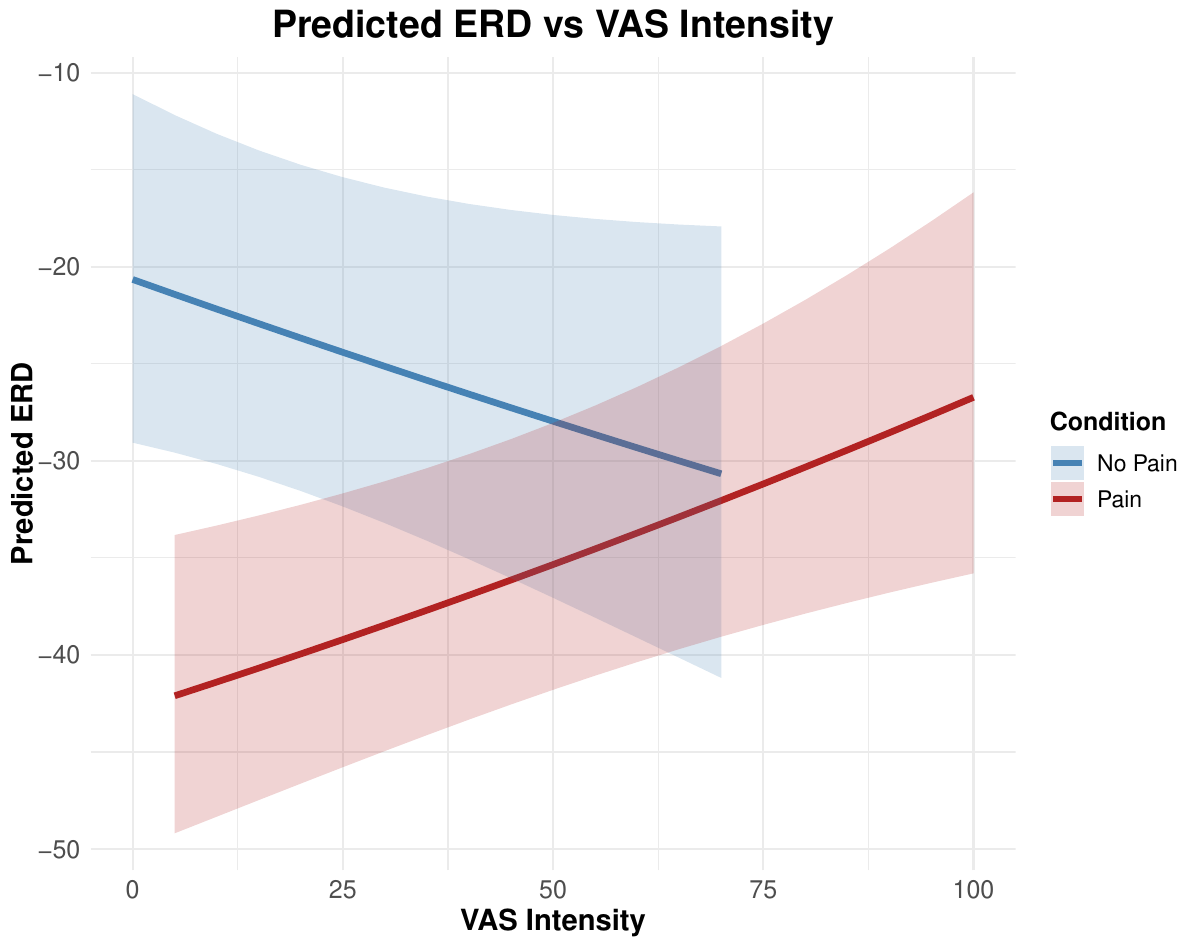}
  \caption{Top-Left: Predicted ERD values by Condition (Pain vs. No-Pain), showing reduced ERD in Pain trials. Top-Right: Grand-average time–frequency maps per Condition, highlighting beta desynchronization during Pain. Bottom-Left: Significant Age × Condition interaction, with older adults showing stronger ERD. Bottom-Right: Condition × VAS interaction, illustrating that higher pain ratings were associated with less negative ERD.}
  \label{fig:ERD_results}
\end{figure*}

\section{Results}

\subsection{Predicting Brain activity from conditions (ERD model)} 
We modeled single-trial ERD values extracted from each significant time--frequency (TF) region of interest (ROI). The results presented here focus on the \textit{Beta} ROI, defined within the time--frequency window [220--760~ms; 15.21--30 ~Hz]. ERD in this ROI was modeled using a full factorial structure including Condition (Pain vs.\ No-Pain), Age group (Young vs.\ Old), Gender (Male vs.\ Female), and VAS ratings, with all possible multi-way interactions between fixed factors. A backward stepwise reduction procedure was then applied, yielding a parsimonious model that retained only the most relevant predictors:

\begin{equation}
\begin{split}
\text{ERD}_{\text{Beta}} \sim {} & \text{Age} + \text{Condition} + \text{Gender} \\
& {} + \text{Age}:\text{Condition} + \text{VAS\_Intensity} \\
& {} + \text{Condition}:\text{VAS\_Intensity} + \text{Condition}:\text{Gender} \\
& {} + (1\,|\,\text{PID})
\end{split}
\end{equation}

\subsubsection{Pain versus No-Pain Conditions}
Across the group, event-related desynchronization (ERD) values in the beta frequency range were significantly reduced during Pain compared to No-Pain trials. Generalized linear mixed models (GLMMs) with Gamma distribution revealed a strong main effect of Condition (see left in \ref{fig:ERD_results}), with lower ERD values under Pain ($p = 7.02 \times 10^{-16}$). This effect was robust across model specifications and was consistent with the expected desynchronization of frontal-central oscillations in response to nociceptive input \cite{Ploner2017, Nickel2017}. 

Notably, the Condition effect was modulated by Age. Older adults show greater ERD (more negative), and the age difference was larger for pain than for no-pain trials. A significant Age $\times$ Condition interaction ($p < 0.001$) confirmed that oscillatory markers of pain are sensitive to age-related differences in neural processing. Gender effects were weaker overall but contributed to model fit when interactions were considered.

\subsubsection{Coupling of ERD with Subjective Intensity (VAS)}
When subjective intensity ratings were incorporated, ERD values scaled with perceived pain. Within the Pain condition, higher VAS scores were associated with less negative ERD (i.e., weaker desynchronization), suggesting that beta activity remained closer to baseline at higher pain levels (Shown in the Right of \ref{fig:ERD_results}). The Condition $\times$ VAS interaction was significant ($p = 0.002$), supporting the link between oscillatory activity and perceived intensity. Gender also moderated these effects, with interaction terms indicating differences in ERD–VAS coupling between male and female participants ($p = 0.02$).

\subsection{Prediction of Pain Intensity from ERD (VAS Model)}
\begin{figure*}
  \centering
  \includegraphics[width=.49\linewidth]{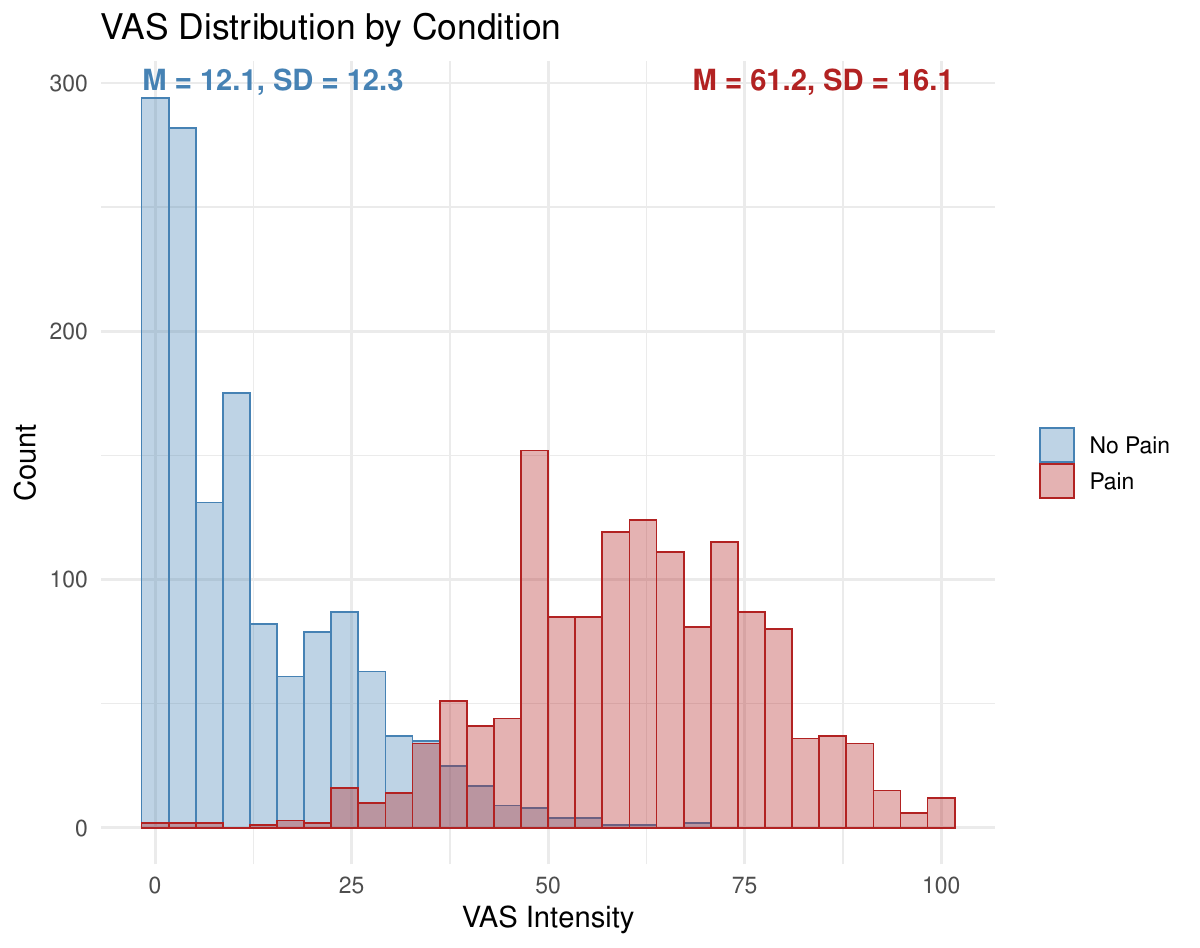}
  \includegraphics[width=.49\linewidth]{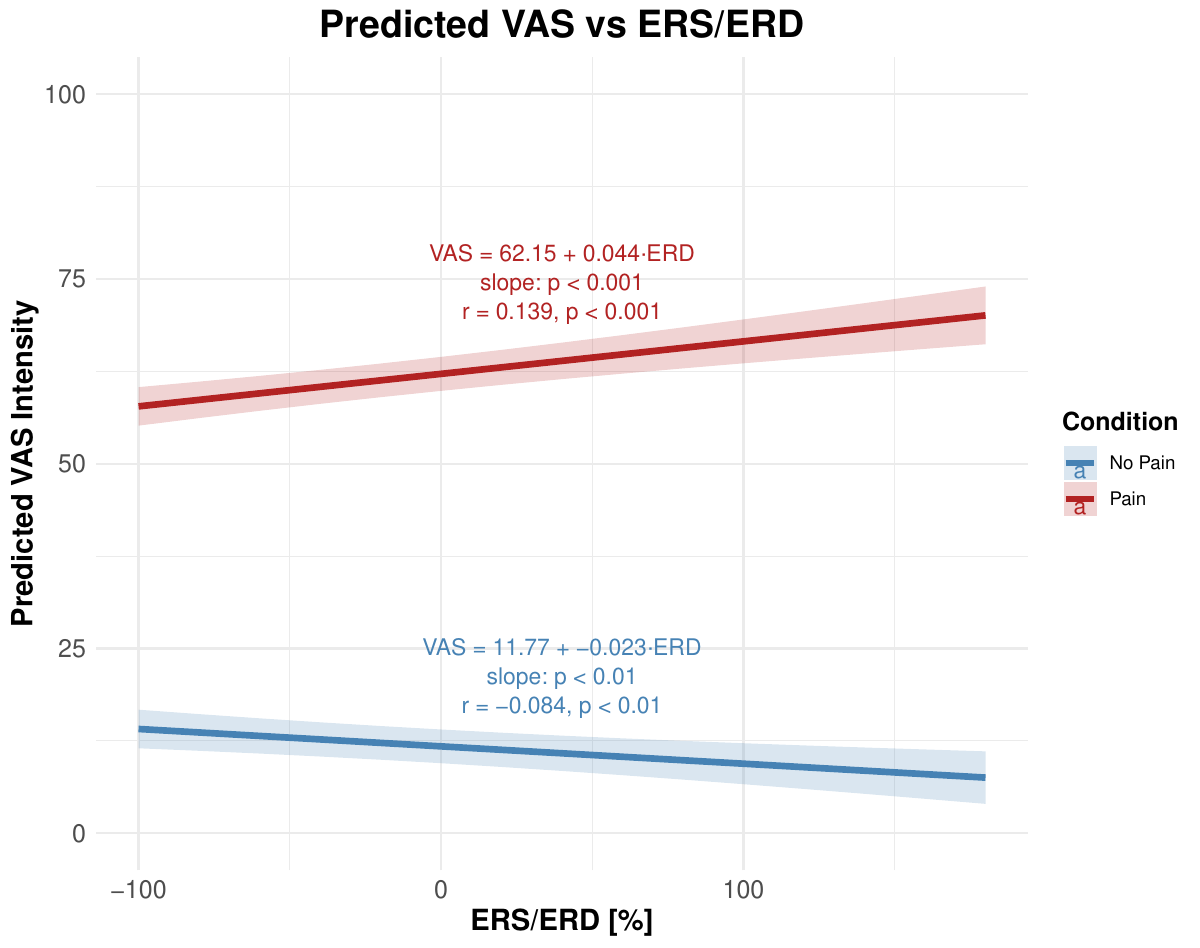}
  \caption{Left: Distribution of VAS intensity ratings across Pain (Red) and No-Pain (Blue) conditions, showing distinct distributional profiles. Right: Predicted VAS values based on ERD and Condition, with regression slopes indicating significant ERD–VAS coupling. Shaded areas represent 95\% confidence intervals.}
  \label{fig:VAS_results}
\end{figure*}

Reverse models were tested to evaluate the predictive capacity of ERD on subjective VAS ratings. ERD alone significantly predicted VAS across individuals ($p = 3.42 \times 10^{-7}$), with stronger effects under the Pain condition. Importantly, the distribution of VAS scores differed between conditions: the No-Pain condition followed a Gamma-like distribution, whereas the Pain condition approximated a Normal distribution (see Figure~\ref{fig:VAS_results}, left panel). This heterogeneity complicates modeling unless Condition is explicitly included. 

To account for this, a second model was tested including the interaction between ERD and Condition. This analysis revealed significant effects of ERD ($p < 1 \times 10^{-3}$), Condition ($p < 2 \times 10^{-16}$), and their interaction ($p = 2 \times 10^{-10}$), see Right part in Figure \ref{fig:VAS_results}). Model performance further improved when demographic variables were added: Age and Gender both contributed to explained variance, particularly in interaction with Condition. Specifically, Condition $\times$ Age ($p = 2 \times 10^{-14}$), Condition $\times$ Gender ($p = 1 \times 10^{-4}$), and the three-way Condition $\times$ Age $\times$ Gender interaction ($p = 2 \times 10^{-6}$) were all significant. Moreover, Age showed a significant interaction with ERD ($p = 4 \times 10^{-2}$). 


\subsection{Key Findings}
In summary, frontal-central ERD robustly distinguished Pain from No-Pain conditions, scaled with subjective intensity ratings, and was moderated by age and gender. These results highlight the potential of trial-level oscillatory markers as individualized neural signatures of pain. Figures~\ref{fig:TF_steps} and \ref{fig:VAS_results} illustrate the grand-average time--frequency maps and GLMM-based predictions of VAS, respectively.

\section{Discussion and Conclusions}

\subsection{Summary of Findings}
The present study provides preliminary evidence that oscillatory brain activity measured with EEG can serve as a reliable neural marker of pain at the single-trial level. Using a high-density EEG dataset from 59 participants, we demonstrated that beta-band event-related desynchronization (ERD) in frontal-central regions (F1, F2, Fz, FCz) robustly distinguished Pain from No-Pain conditions. Importantly, ERD values scaled with subjective pain intensity ratings and were further modulated by age and gender, suggesting individualized neurophysiological signatures of pain perception. Reverse modeling confirmed that ERD could significantly predict subjective intensity across individuals, supporting its potential as a nonverbal index of pain. Together, these findings advance the field by moving beyond trial-averaged analyses, establishing the feasibility of per-trial oscillatory decoding, and identifying demographic moderators of pain-related brain activity.  

Our modeling results also suggested demographic modulation of oscillatory markers: 
for example, predicted ERD values were higher in young males compared to young females reporting the same intensity level. 
This pattern supports the notion of individualized neural fingerprints of pain perception.

\subsection{Interpretation in the Context of Literature}
Our results are consistent with prior work linking nociceptive processing to desynchronization in the alpha and beta frequency ranges across sensory and frontal cortices \cite{Ploner2017, Nickel2017, Peng2014, Hu2019}. The observed reduction in ERD during Pain relative to No-Pain conditions aligns with studies suggesting that desynchronization reflects cortical engagement in nociceptive processing \cite{Dowman2008}. 

The coupling of ERD with subjective intensity ratings also supports the notion that oscillatory activity provides a dynamic neural correlate of perceived pain. Interestingly, we observed that higher VAS ratings were associated with less negative ERD, contrary to the intuitive expectation that stronger pain would elicit stronger desynchronization. One possible explanation is that increased Beta power (less negative ERD) at higher pain levels, could reflect compensatory processes such as increased cognitive control or attentional engagement in response to high-intensity pain. Similar patterns have been reported in pain studies, where oscillatory dynamics do not always scale monotonically with reported intensity. For example, Nickel et al. (2017) \cite{Nickel2017} and Hu \& Iannetti (2019) \cite{Hu2019}  showed that alpha/beta oscillations can dissociate stimulus intensity from perceived pain. Early work by Dowman et al (2008) \cite{Dowman2008}  also reported complex tonic-pain ERD/ERS dynamics. Conceptually, these paradoxical relationships are consistent with reviews highlighting that oscillatory markers capture not only sensory encoding but also top-down modulation, such as attention and expectancy effects \cite{Ploner2017}. Future work is needed to disentangle whether this counterintuitive ERD--VAS coupling reflects neural ceiling effects, compensatory recruitment, or methodological aspects of baseline normalization.

The moderating role of age and gender observed here extends previous findings on age-related decline in pain modulation \cite{vanderMeulen2024} and gender differences in pain sensitivity. These results highlight the importance of incorporating demographic factors when evaluating neural signatures of pain, as they may account for interindividual variability often overlooked in trial-averaged analyses.  

\subsection{Potential Impact}
The ability to decode pain conditions at the single-trial level has significant implications for both basic neuroscience and clinical translation. From a scientific perspective, these results contribute to a deeper understanding of the temporal dynamics of pain perception and the role of demographic variables in shaping oscillatory responses. From a translational perspective, establishing EEG-derived ERD as a nonverbal marker of pain could provide objective tools for monitoring pain in non-communicative patients, such as those under anesthesia, in intensive care, or with impaired cognitive function. Furthermore, such markers could serve as endpoints in the evaluation of analgesics or neuromodulatory interventions, providing higher sensitivity and reproducibility compared to self-reports alone.  

\subsection{Future Directions}
While promising, these findings represent an initial step toward objective pain monitoring. Future work should aim to validate these results in independent cohorts and extend analyses to patient populations suffering from chronic or neuropathic pain. Methodologically, future research could expand beyond frontal-central electrodes to include network-level analyses, connectivity measures, and multiband oscillatory interactions. Integrating EEG with structural and functional MRI would further improve spatial resolution and enable the identification of multimodal "pain phenotypes".  

In addition, the predictive models developed here could be extended using machine learning and explainable AI approaches to uncover distributed spatiotemporal features with greater predictive power, using explainable tools \cite{Gillet2025} that can facilitate the comprehension of results. Another important avenue is to investigate whether externally induced synchronization, such as sustained gaze fixation or rhythmic visual flicker, can counteract pain-related desynchronization, offering mechanistically grounded and non-invasive strategies for pain modulation \cite{Ecsy2018, LopezDiaz2021}.  

\subsection{Concluding Remarks}
In conclusion, this study demonstrates that EEG-derived ERD in the beta band provides a reproducible and individualized neural signature of pain perception. By showing that these oscillatory markers distinguish pain conditions, scale with subjective intensity, and vary across demographic groups, our findings open new avenues for objective, report-free pain monitoring. These results lay the groundwork for future multimodal and intervention-based approaches that may ultimately contribute to more personalized and patient-friendly pain assessment and treatment strategies.  




\begin{thebibliography}{99}

\bibitem{Ploner2017}
M.~Ploner, C.~Sorg, and J.~Gross, ``Brain rhythms of pain,'' \emph{Trends in Cognitive Sciences}, vol.~21, no.~2, pp. 100--110, 2017.

\bibitem{Peng2014}
W.~Peng, L.~Hu, Z.~Zhang, and Y.~Hu, ``Changes of spontaneous oscillatory activity to tonic heat pain,'' \emph{PLoS One}, vol.~9, no.~3, p. e91052, 2014.

\bibitem{Nickel2017}
M.~M. Nickel, E.~S. May, L.~Tiemann, P.~Schmidt, M.~Postorino, S.~Ta~Dinh, F.~Nees, H.~Flor, and M.~Ploner, ``Brain oscillations differentially encode noxious stimulus intensity and pain intensity,'' \emph{NeuroImage}, vol. 148, pp. 141--147, 2017.

\bibitem{Hu2019}
L.~Hu and G.~D. Iannetti, ``Neural indicators of perceptual variability of pain across species,'' \emph{Proceedings of the National Academy of Sciences}, vol. 116, no.~5, pp. 1782--1791, 2019.

\bibitem{Dowman2008}
R.~Dowman, D.~Rissacher, and S.~Schuckers, ``{EEG} indices of tonic pain-related activity in the somatosensory cortices,'' \emph{Clinical Neurophysiology}, vol. 119, no.~5, pp. 1201--1212, 2008.

\bibitem{Diers2020}
M.~Diers, J.~Koeppe, N.~Diesch, \emph{et~al.}, ``Induced oscillatory signaling in the beta frequency of top-down pain modulation,'' \emph{Pain Reports}, vol.~5, no.~1, p. e806, 2020.

\bibitem{Ecsy2018}
K.~Ecsy, C.~A. Brown, and A.~K.~P. Jones, ``Cortical nociceptive processes are reduced by visual alpha-band entrainment in the human brain,'' \emph{European Journal of Pain}, vol.~22, no.~3, pp. 538--550, 2018.

\bibitem{vanderMeulen2024}
M.~van~der Meulen, K.~M. Rischer, A.~M. Gonz{\'a}lez-Rold{\'a}n, J.~L. Terrasa, P.~Montoya, and F.~Anton, ``Age-related differences in functional connectivity associated with pain modulation,'' \emph{Neurobiology of Aging}, vol. 140, pp. 1--11, 2024.

\bibitem{Veniero2021}
D.~Veniero, J.~Gross, S.~Morand, F.~Duecker, A.~T. Sack, and G.~Thut, ``Top-down control of visual cortex by the frontal eye fields through oscillatory realignment,'' \emph{Nature Communications}, vol.~12, p. 1691, 2021.

\bibitem{Quentin2016}
R.~Quentin, F.~Chanes, J.~Vernet, \emph{et~al.}, ``Visual contrast sensitivity improvement by right frontal high-beta activity is mediated by contrast gain mechanisms and influenced by fronto-parietal white matter microstructure,'' \emph{Cerebral Cortex}, vol.~26, no.~6, pp. 2381--2390, 2016.

\bibitem{LopezDiaz2021}
K.~L{\'o}pez-D{\'\i}az, J.~Henshaw, A.~J. Casson, C.~A. Brown, J.~R. Taylor, N.~J. Trujillo-Barreto, L.~J. Arendsen, and A.~K.~P. Jones, ``Alpha entrainment drives pain relief using visual stimulation in a sample of chronic pain patients: A proof-of-concept controlled study,'' \emph{NeuroReport}, vol.~32, no.~5, pp. 394--398, 2021.

\bibitem{vanHoornweder2022a}
S.~van Hoornweder, D.~A. Blanco{-}Mora, S.~Depestele, K.~van Dun, K.~Cuypers, S.~Verstraelen, and R.~Meesen, ``Aging and complexity effects on hemisphere-dependent movement-related beta desynchronization during bimanual motor planning and execution,'' \emph{Brain Sciences}, vol.~12, no.~11, p. 1444, 2022.

\bibitem{vanHoornweder2022b}
S.~van Hoornweder, D.~A.~B. Blanco{-}Mora, S.~Depestele, J.~Frieske, K.~van Dun, K.~Cuypers, S.~Verstraelen, and R.~Meesen, ``Age and interlimb coordination complexity modulate oscillatory spectral dynamics and large-scale functional connectivity,'' \emph{Neuroscience}, vol. 496, pp. 1--15, 2022.

\bibitem{BlancoMora2024}
D.~A. Blanco{-}Mora, S.~van Hoornweder, K.~van Dun, S.~Verstraelen, K.~Cuypers, S.~Berm{\'u}dez~i Badia, and R.~Meesen, ``Toward methodologies for motor imagery enhancement: a t{DCS}-{BCI} study,'' \emph{Brain-Computer Interfaces}, vol.~11, no.~3, pp. 110--124, 2024.

\bibitem{vanHoornweder2025}
S.~van Hoornweder, D.~A. Blanco{-}Mora, M.~Nuyts, K.~Cuypers, S.~Verstraelen, and R.~Meesen, ``The causal role of beta band desynchronization: Individualized high-definition transcranial alternating current stimulation improves bimanual motor control,'' \emph{NeuroImage}, vol. 312, p. 121222, 2025.

\bibitem{Rischer2025}
K.~M. Rischer, A.~M. Dierolf, F.~Anton, P.~Montoya, A.~M. Gonz{\'a}lez{-}Rold{\'a}n, and M.~van~der Meulen, ``The placebo hypoalgesic response is reduced in healthy older adults showing a decline in executive functioning,'' \emph{Journal of Pain Research}, vol.~18, pp. 1747--1763, 2025.

\bibitem{Delorme2004}
A.~Delorme and S.~Makeig, ``{EEGLAB}: an open source toolbox for analysis of single-trial {EEG} dynamics including independent component analysis,'' \emph{Journal of Neuroscience Methods}, vol. 134, no.~1, pp. 9--21, 2004.

\bibitem{Gustafsson1996}
F.~Gustafsson, ``Determining the initial states in forward-backward filtering,'' \emph{IEEE Transactions on Signal Processing}, vol.~44, no.~4, pp. 988--992, 1996.

\bibitem{PionTonachini2019}
L.~Pion{-}Tonachini, K.~Kreutz{-}Delgado, and S.~Makeig, ``{ICLabel}: an automated electroencephalographic independent component classifier, dataset, and website,'' \emph{NeuroImage}, vol. 198, pp. 181--197, 2019.

\bibitem{Cohen2014}
M.~X. Cohen, \emph{Analyzing Neural Time Series Data: Theory and Practice}.\hskip 1em plus 0.5em minus 0.4em\relax Cambridge, MA: MIT Press, 2014.

\bibitem{Bates2015}
D.~Bates, M.~M{\"a}chler, B.~Bolker, and S.~Walker, ``Fitting linear mixed-effects models using lme4,'' \emph{Journal of Statistical Software}, vol.~67, no.~1, pp. 1--48, 2015.

\bibitem{RStudio2020}
{RStudio Team}, ``{RStudio}: Integrated development for {R},'' \url{http://www.rstudio.com/}, 2020, rStudio, PBC, Boston, MA.

\bibitem{Gillet2025}
T.~Gillet, M.~Oliveira, B.~Bausch, D.~A. Blanco{-}Mora, J.~Gon{\c{c}}alves, E.~Blaney~Davidson, and M.~van~der Meulen, ``G3d-vit, a small step towards 3d interpretability of vision transformers,'' in \emph{Proceedings of the 11th World Congress on Electrical Engineering and Computer Systems and Sciences}, 2025, mVML 138.

\end{thebibliography}




\end{document}